\documentclass[aps,rmp,reprint,a4paper,showpacs,showkeys,floats,onecolumn,nofootinbib,notitlepage,hyperref]{revtex4-1}
\usepackage{amsfonts}
\usepackage{amsmath}
\usepackage{amssymb}
\usepackage{graphicx}%

\begin{document}

\preprint{Phys. Rev. E {\bf 80}, 062101 (2009)}

\title{Effectiveness of the Kozachenko-Leonenko estimator for generalised entropic forms
\thanks{Published reference: Phys. Rev. E {\bf 80}, 062101 (2009)}
}
\author{S\'{\i}lvio M. Duarte Queir\'{o}s}
\affiliation{Unilever R\&D Port Sunlight, Quarry Road East, Wirral, CH63 3JW UK}
\email{Silvio.Queiros@unilever.com,sdqueiro@gmail.com}

\begin{abstract}
In this manuscript we discuss the effectiveness of the Kozachenko-Leonenko entropy estimator when generalised to cope with
entropic forms customarily applied to study systems evincing asymptotic scale invariance and dependence (either linear or non-linear type).
We show that when the variables are independently and identically distributed the estimator is only valuable along the whole domain if the data follow the uniform distribution, whereas for other distributions the estimator is only effectual in the limit of the Boltzmann-Gibbs-Shanon entropic form. We also analyse the influence of the dependence (linear and non-linear) between variables on the accuracy of the estimator between variables. As expected in the last case the estimator looses efficiency for the Boltzmann-Gibbs-Shannon entropic form as well.
\end{abstract}

\date{16th June 2009}

\maketitle

\tableofcontents

\section{Introduction}

After a period of harsh criticism, the connection between the microscopic world and the displayed macroscopic
properties of the system by means of the Boltzmann principle, which was later extended by
Gibbs to systems in contact with a reservoir, has achieved an incontestable consensus~[\cite{cohen}].
Despite its broad acceptance, it is neglected by many people that the
standard statistical mechanics is still based on a hypothesis, the \emph{Stosszahl
Ansatz}~[\cite{huang}]. This ansatz is intimately related to the ergodic theory which
has only been analytically proven for a set of very few simple systems~[\cite{sinai}].
With the surging interest in more intricate systems for which the ergodic theory is bound to be invalid,
\textit{e.g.}, systems that occupy their allowed phase space in a
scale-invariant way or exhibit long spatiotemporal correlations~[\cite{tsallis-pnas}],
entropic forms different to the Boltzmann-Gibbs (BG) functional have been presented. Among several,
two of them might be given special emphasis: the Renyi entropy~[\cite{renyi}] and the non-additive
entropy proposed in a physical context by Tsallis~[\cite{tsallis}]. For the last two decades there has been an
impressive amount of work towards the physical validation and application of
the latter~[\cite{book}]. As occurs in the BG standard case~[\cite{abramov,swinney,klakl}],
many systems studied within the non-additive formalism present a reduced number of observations or correspond to finite
size systems~[\cite{ct-caruso,phys-d}]. Consequently, a considerable error can be introduced if the simplest method based on binning
the data is assumed and the number of observables is very small.

In this manuscript we generalise a well-known binless strategy for the
estimation of BG entropy, the Kozachenko-Leonenko algorithm (KLA)~[\cite{kla}],
which bases the estimation of the theoretical entropy on the distance $\delta / 2$
to the nearest-neighbour of a specific order $n$. We illustrate its possible validity by
comparing numerical results with the theoretical values in two different situations:
independent and dependent variables. In the former, we survey three standard
distributions (PDF), namely the Gaussian, the Student-$t$
(or $q$-Gaussian) and the uniform PDF. In the latter case, we analyse linear and
non-linear dependent Student-$t$ variables. For the sake of simplicity, we will
restrict our analysis to one-dimensional systems corresponding to sets of random variables.

\section{Generalising KLA}

The non-additive entropy is defined as~[\cite{tsallis}],%
\begin{equation}
S_{Q}\equiv\frac{1-\int\left[  p\left(  x\right)  \right]  ^{Q}\,dx}%
{Q-1},\qquad\left(  Q\in\mathbb{R}\right)  \label{s-q}%
\end{equation}
which in the limit $Q$ going to $1$ concurs with the Boltzmann-Gibbs entropy,
$S_{1}=S_{BG}\equiv-\int p\left(  x\right)  \,\ln\,p\left(  x\right)
\,dx=-\left\langle \ln\,p\left(  x\right)  \right\rangle $, where
$\left\langle \ldots\right\rangle $ represents the average. Bearing in mind the
$Q$-logarithm definition
\cite{book},
$\lim_{Q\rightarrow1}\left\{  \ln_{Q}x\equiv\frac{x^{1-Q}-1}{1-Q}\right\}
=\ln\,x$,
it is easily
verifiable that the entropic functional can be written in the following way,%
\begin{equation}
S_{Q}^{\ast}=-\int p\left(  x\right)  \,\ln_{q}\,p\left(  x\right)
\,dx=-\left\langle \ln_{q}\,p\left(  x\right)  \right\rangle
,\label{2-q_surprise}%
\end{equation}
with $Q=2-q$. In other words, the entropy $S_{Q}$ represents the average value
of an alternative way of describing the \textit{surprise}. From this definition, we are
evoked to apply the same ideas of the binless KLA.

Let us consider a set of $N$ random variables, $\left\{  x_{i}\right\}  $,
identically distributed and associated with a generic PDF,
$p\left(  x\right)$, whose entropy estimation works out at,%
\begin{equation}
S_{Q}=-\frac{1}{N}\sum_{i}\ln_{q}\,P\left(  x_{i}\right) \equiv - \left\langle \ln_{q} \, P_i \right\rangle
,\label{sigma-estimator}%
\end{equation}
where $P\left(  x\right)  \approx\delta\,p\left(  x\right)  $ (here $\delta$
represents a segment of the $x$ domain which preferentially tends to 0). Equation~(\ref{sigma-estimator}) should be
equal to $S_{Q}^{\ast}$ in the limit of $N$ going to infinity and
$\delta\rightarrow0$. Alternatively, the measure $P\left(  x_i \right) $ relates to the
distance $\delta $ (centered at $x_i$) which comprises a given number of nearest-neighbours, $n$ (originally $n=1$), or accordingly to the
probability $\Pi_{n}\left(  \delta\right)$ that the $\left(  n-1\right)$ nearest-neighbours have values
$x^{\prime}$ within $x\pm\delta/2$ and the $n$-th nearest
neighbour is at a distance $\delta/2$ of $x_i$, \textit{i.e.},
\begin{equation}
\Pi_{n}\left(  \delta\right)  =\frac{\left(  N-1\right)  !}{\left(
n-1\right)  !\left(  N-n-1\right)  !}\frac{\left[  P_{i}^{\prime}\left(
\delta\right)  \right]  ^{n-1}}{\left[  1-P_{i}^{\prime}\left(  \delta\right)
\right]  ^{1+n-N}}\frac{d\,P_{i}^{\prime}\left(  \delta\right)  }{d\,\delta
},\label{prob-n-delta}%
\end{equation}
where $P_{i}^{\prime}\left(  \delta\right)  =\int_{x-\delta/2}^{x+\delta
/2}p_{i}^{\prime}\left(  z\right)  \,dz$. Thence, we associate $\left\langle \ln_{q} \, P_i \right\rangle$
with $ \overline{\ln_{q}P_{i}^{\prime}} = \int$ $\Pi_{n}\left(  \delta\right)  \ln_{q}\,P_{i}^{\prime
}\left(  \delta\right)  \,d\delta$ that yields~[\cite{gradshteyn}],
\begin{equation}
\overline{\ln_{q}P_{i}^{\prime}}=\frac{1-\frac{\Gamma\left[  N\right]
\Gamma\left[  n+1-q\right]  }{\Gamma\left[  n\right]  \Gamma\left[
1+N-q\right]  }}{q-1}.
\label{probbarra}
\end{equation}
Taking into consideration that $\ln_{q}(u \times v)=\ln_{q}u+\ln_{q}v+\left(
1-q\right)  \ln_{q}u \times \ln_{q}v$ and remembering that $P_{i}^{\prime}\left(
\delta\right)  \approx p_{i}^{\prime}\,\delta$ we obtain the final
formula,%
\begin{equation}
S_{Q}=\frac{\overline{\ln_{q}P_{i}^{\prime}}-\left\langle \ln_{q}%
\,\delta\right\rangle }{1+\left(  1-q\right)  \left\langle \ln_{q}%
\,\delta\right\rangle },\label{sq-algoritmo}%
\end{equation}
where $\left\langle \ln_{q}\,\delta\right\rangle $ represents the average of $\ln_{q}\,\delta$ over all points and samples accessible.

In practical terms, the algorithm is implemented the following way. For a fixed order of the vicinity, the distance $\delta / 2$
from each point $x_i$ of the dataset under study to its $n$-th nearest neighbour is determined.
The values of $\delta $ are then used to compute the average of $\ln_q \delta$ that is used in the previous equation.
The value of $\overline{\ln_{q}P_{i}^{\prime}}$ is pre-defined when the values of $q$ and $n$ used in Eq.~(\ref{probbarra}) are fixed.

Endowed with Eq.~(\ref{sq-algoritmo}), we can rate the quality of the
approximation by comparing its outcome with the predicted theoretical values given by
Eq.~(\ref{2-q_surprise}). For the cases we will present hereinafter we have,%
\begin{equation}
S_{2-q}^{\ast}=\frac{1}{1-q}-\frac{\left(  2\,\pi\right)  ^{\frac{q-1}{2}}%
}{\left(  1-q\right)  \sqrt{2-q}},\label{gaussian}%
\end{equation}
for the Gaussian,%
\begin{equation}
S_{2-q}^{\ast}=\frac{1}{1-q}-\frac{2^{2-q}\,3^{\frac{q-1}{2}}\Gamma\left[
\frac{7}{2}-2\,q\right]  }{\pi^{\frac{3-q}{2}}\left(  1-q\right)
\,\Gamma\left[  4-2\,q\right]  },\label{student-t}%
\end{equation}
for the Student-$t$ with $3$ degrees of freedom\footnote{Because, under appropriate constraints, the entropy
$S_{Q}$ is maximised by the Student-$t$ PDF, the latter has been also named $Q$-Gaussian distribution wherein the
relation $Q=\frac{3+m}{1+m}$ between the entropic index, $Q$, and the degree
of freedom $m$ is valid~[\cite{andre}].} and%
\begin{equation}
S_{2-q}^{\ast}=-\ln_{q}\,2,\label{uniform}%
\end{equation}
for a uniform PDF defined between $-1$ and $1$.

\section{Results}

\begin{figure}[tbh]
\begin{center}
\includegraphics[width=0.6\columnwidth,angle=0]{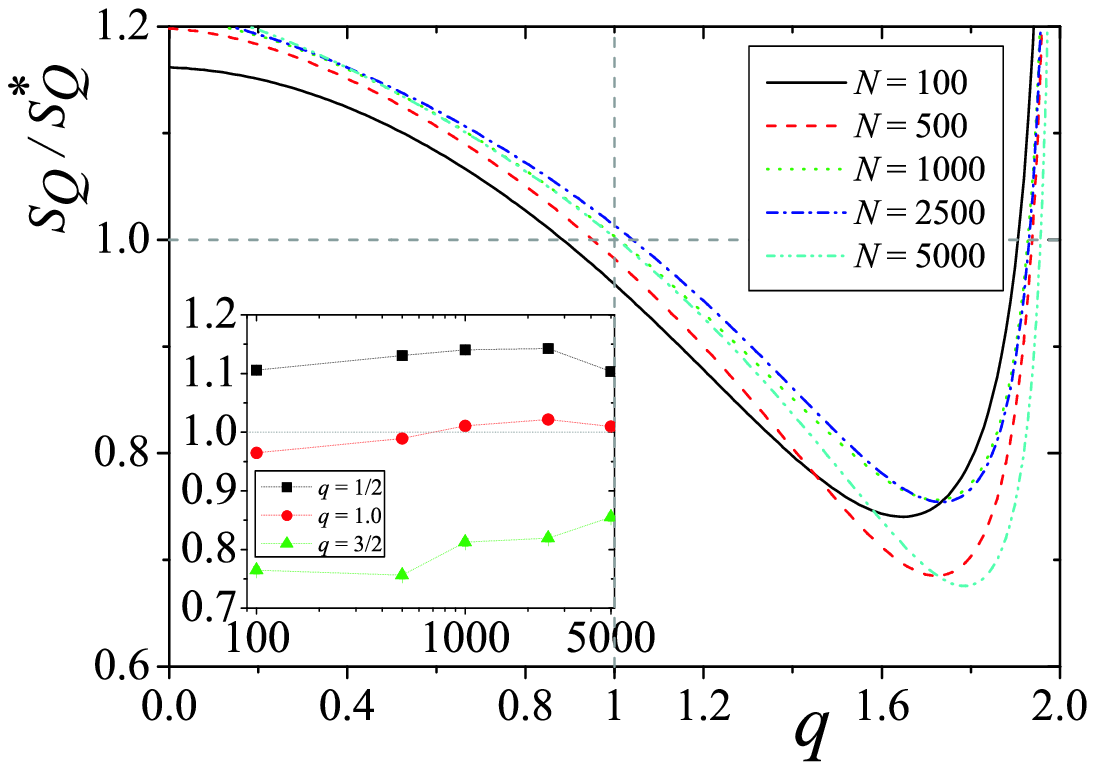}
\end{center}
\caption{(Colour online) Ratio $S_{Q} / S_{Q}^{\ast}$ vs the dual entropic parameter $q=2-Q$
for fixed $n=1$. The inset depicts the same ratio vs $N$ for particular values of $q$.
In this case the sets are composed of Gaussian distributed random variables.}%
\label{fig-1}%
\end{figure}

\begin{figure}[tbh]
\begin{center}
\includegraphics[width=0.6\columnwidth,angle=0]{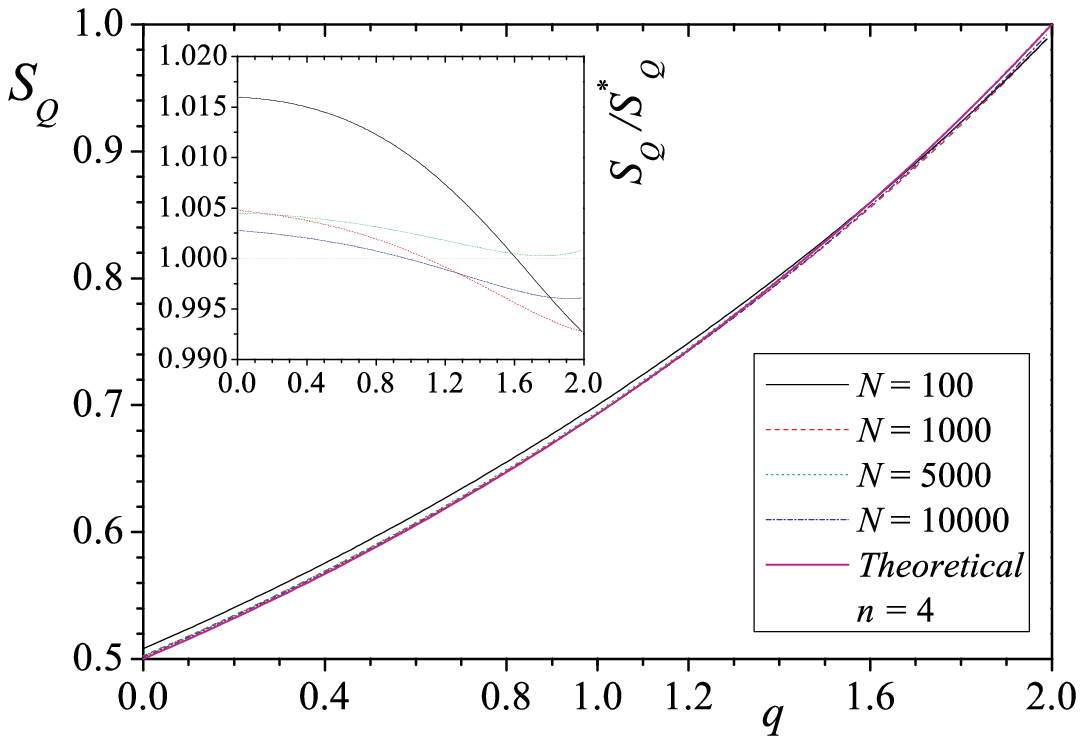}
\includegraphics[width=0.6\columnwidth,angle=0]{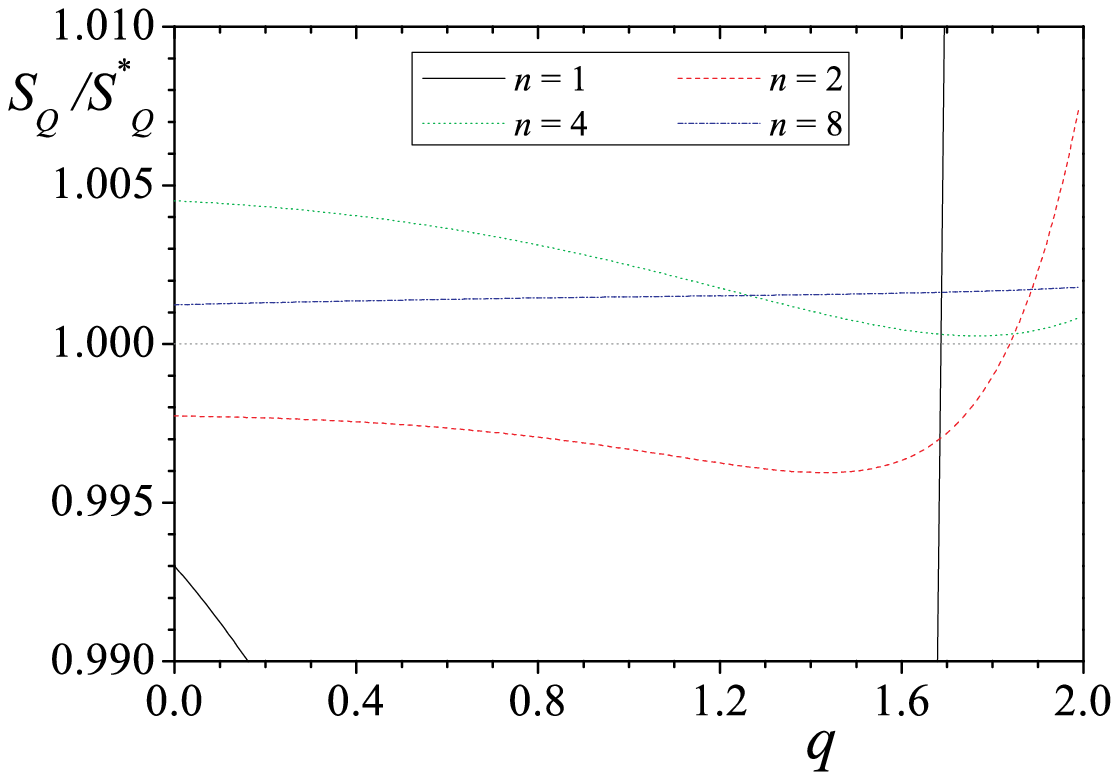}
\end{center}
\caption{(Colour online) Upper panel:
Estimated $S_{Q}$ vs the dual entropic parameter $q=2-Q$ [the inset represents the ratio
$S_{Q}/S_{Q}^{\ast}$ vs $q$]; Lower panel: Ratio $S_{Q} / S_{Q}^{\ast}$ vs $q$ for different
$n$ $N=5000$. In this case the sets are composed of uniformly
distributed random variables between -1 and 1 with the number of samples taken
into account referred in the text.}%
\label{fig-3}%
\end{figure}

In order to test the actual efficiency of Eq.~(\ref{sq-algoritmo}) we
generated sets (typically $10^{3}$) of random variables with a number of
elements never larger than $10^{4}$ on which we have applied the algorithm for
diverse values of $n$.\footnote{The random variables were bore by means of the
Extended Cellular Automata random number generator using the five-neighbour
rule [\cite{gentle}]. Additionally for the case of the Student-$t$ we used the
Bailey algorithm~[\cite{bailey}].} The results depicted in Figs. \ref{fig-1}%
-\ref{fig-3} show that for the Gaussian and the Student-$t$, the
Kozachenko-Leonenko approach is only a valuable estimator for values of $q=1$,
\textit{i.e.}, for the BG case, whereas for the uniform PDF it is quite effective.

For the Gaussian (see Fig.~\ref{fig-1}), we have verified that for
$N<5000$ we have got error greater than $10\%$ unless we are analysing the
$q=Q=1$ value. In this case, the error is already less than $1\%$ for $N=100$.
For the remaining $q\neq1$ cases, we have not captured a monotonous behaviour of
the error and the ratio $S_{Q}/S_{Q}^{\ast}$ with the number of elements of the
set or the order of the neighbour used. In respect of the dependence of $S_{Q}/S_{Q}^{\ast}$ on $n$ (for fixed $N$),
we have verified alike behaviour with $n=1$ which presents the best estimations for any fixed $N$ tested.

\begin{figure}[tbh]
\begin{center}
\includegraphics[width=0.6\columnwidth,angle=0]{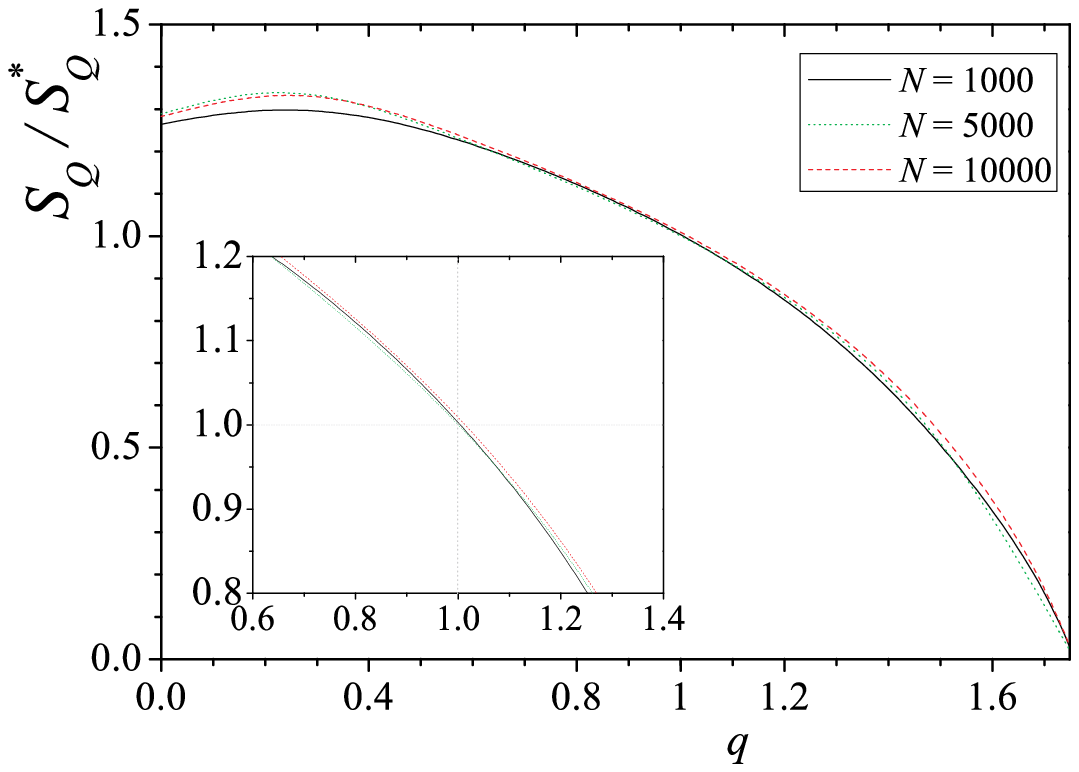}
\includegraphics[width=0.6\columnwidth,angle=0]{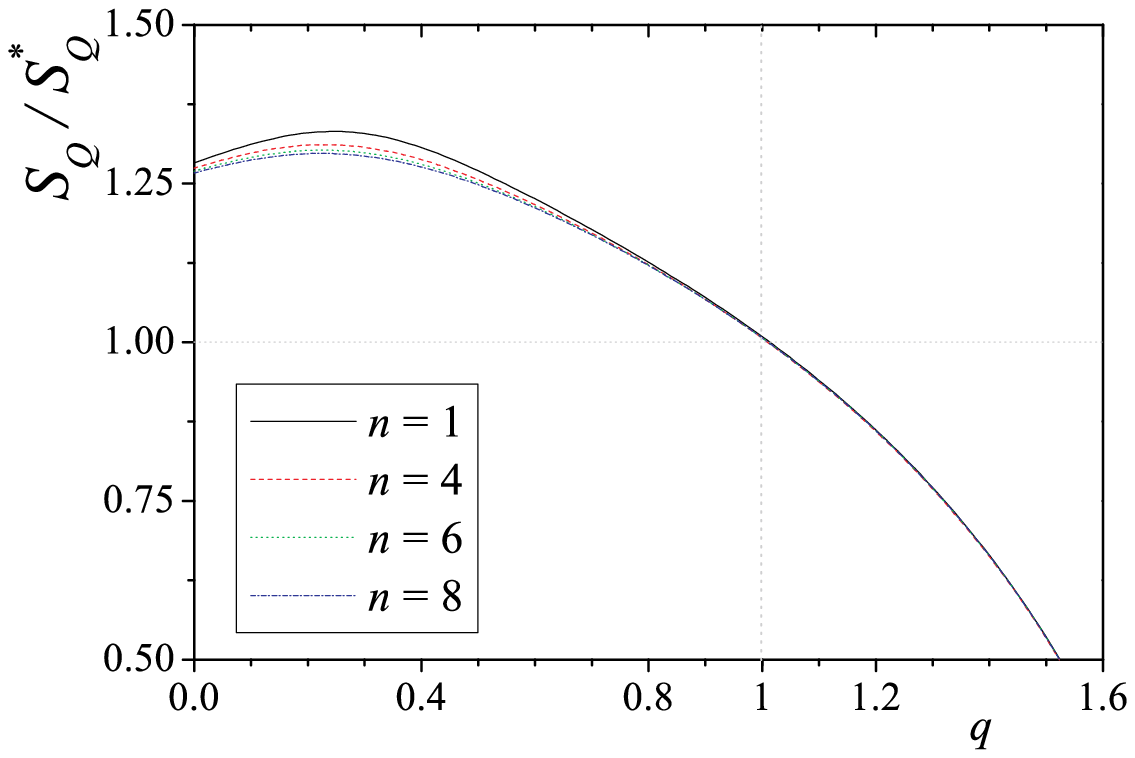}
\end{center}
\caption{(Colour online) Upper panel: Ratio $S_{Q} / S_{Q}^{\ast}$ vs the dual entropic parameter $q=2-Q$ for
fixed $n=1$; Lower panel: The same but for different $n$ and fixed $N=5000$.
In this case the sets are composed of Student-$t$ (with $3$ degrees
of freedom) distributed random variables.}%
\label{fig-2}%
\end{figure}

Regarding the Student-$t$ case, we have noticed the same qualitative results,
\textit{i.e.}, the KLA algorithm tends to overestimate (underestimate) the
entropy $S_{Q}^{\ast}$ for $Q>1$ ($Q<1$) independently of the size of the
series and the order of the nearest-neighbour taken into reference.\footnote{Although only $n=1$ is shown
herein, we let $n$ run up to the remote value of $n=100$.} Once again, for the case $Q=1$ the algorithm caters for an excellent
approach even for relatively small sets ($N<1000$) as exhibited in Fig.~\ref{fig-2}.

As shown in Fig.~\ref{fig-3}, the ineffectiveness we have reported so far is only challenged when the
uniform PDF is considered. In this case, for values of $n>1$, we have verified that the KLA is a trustworthy estimator of
the theoretical entropy of a system. For instance, by considering sets of
$100$ variables we have achieved discrepancies never greater than 2\%. Comparing
the KLA results with entropy evaluations obtained by a simple binning of the
sets we verify the algorithm is only slightly better than the latter approach.
Taking into account the computation time we would say that the KLA does not
pay off.

\begin{figure}[tbh]
\begin{center}
\includegraphics[width=0.6\columnwidth,angle=0]{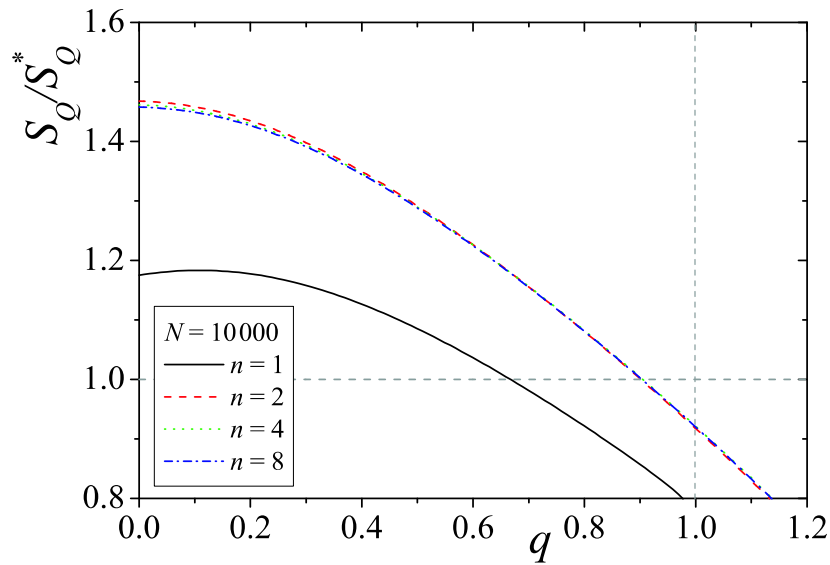}
\end{center}
\caption{(Colour online) Ratio $S_{Q} / S_{Q}^{\ast}$ vs the dual entropic parameter $q=2-Q$
for fixed $N=10000$. In this case the sets are composed of stochastic Feller-like process
as described in the text.}%
\label{fig-4}%
\end{figure}

Complementary, we now study the effectuation of the KLA to time series generated in two different ways (see appendices).
First, we consider the stochastic differential equation
$dx=-\gamma \, x \, dt+ \sqrt{\theta }\, [P(x)]^{\nu } \, dW_{t}$ (It\^{o} notation)~[\cite{borland-pre}] whose stationary PDF is the $q$-Gaussian.
Additionally, the process can reproduce at the first level the intra-day dynamics of the price fluctuations of some financial markets.
We have used $\gamma = 100^{-1}$, $\theta = \gamma \sqrt{2 / \pi }$ and $\nu = -1/2$ which yields the $m =3$ Student-$t$ [($q=1.5$)-Gaussian]
as the stationary PDF. This case is marked by the existence of linear correlations between the variables which affect the quality of the estimation
as plotted in Fig.~\ref{fig-4}. Despite the fact that the best estimative is still for values of $q$ close to 1, the KLA is not so accurate as in
the independent case. Nevertheless, we can surmount this situation taking into consideration that a shuffling procedure does not alter the stationary
PDF of stationary process.

\begin{figure}[tbh]
\begin{center}
\includegraphics[width=0.6\columnwidth,angle=0]{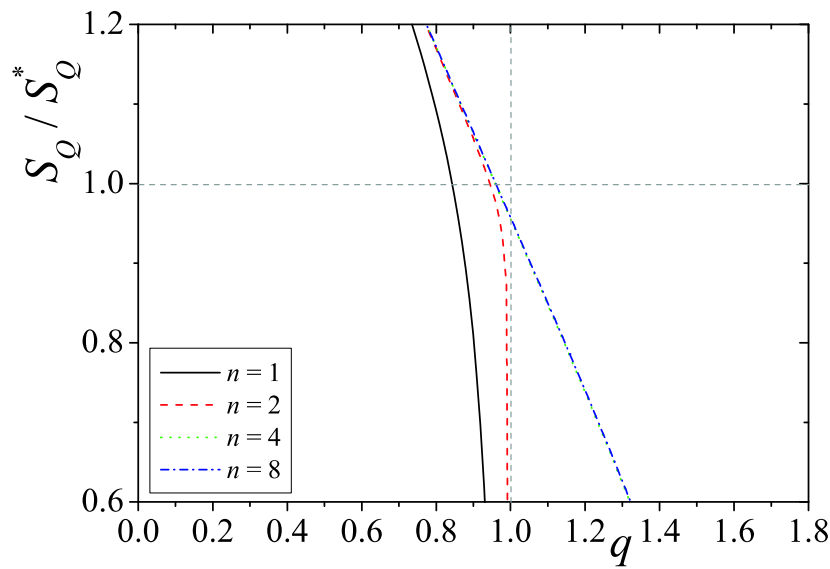}
\end{center}
\caption{(Colour online) Upper panel:
Estimated $S_{Q}$ vs the dual entropic parameter $q=2-Q$ for fixed $N=10000$.
In this case the sets are composed of ($q=1.54$)-Gaussian (approximately) generated
according to the heteroskedastic process described in the text.}%
\label{fig-5}%
\end{figure}

The second case corresponds to time series generated by a heteroskedastic process enclosed within the fractional
ARCH class in which discrete stochastic variables $x_t = \sigma _t \, \omega _t$ ($\omega _t$ follows a Gaussian) are generated with
$\sigma _{t}^{2}=a+b\sum\limits_{i=t_{0}}^{t-1}\mathcal{K}\left( i-t+1\right)
\,x_{i}^{2},  $ where
$\mathcal{K}\left( t^{\prime }\right) \sim \exp _{\zeta} \left[ t^{\prime } \right]$ $\left(
t^{\prime }\leq 0,T>0\right)$~[\cite{q-archepl}] and $\exp _\zeta (\ldots)$ is the inverse function of $\ln _\zeta (\ldots)$.
In spite of generating uncorrelated variables, this model exhibits long-lasting correlations in the variance (non-linear
dependence for $x$) and its probabilistic analysis provides strong statistical evidence that the stationary PDF is a Student-$t$.
Using $\zeta = 1.375$, $b = 0.9375$ and $a=1-b$ we have obtained a ($q=1.54$)-Gaussian. Employing the KLA algorithm, we
have obtained equivalent results to the previous linearly-correlated case (see Fig.~\ref{fig-5}). We must be careful and mind the fact
that the resulting PDF is not exact though. It should be noted that the error in the entropy estimation
is greater than the error presented in the adjustment by a $q$-Gaussian.

\section{Remarks}

In this manuscript we have introduced a generalisation of the well-known
binless Lozachenko-Leonenko entropy estimator to appraise the
(Tsallis) non-additive entropy in systems with a small number of observations
for which binning strategies are likely to present strong deviation from the
expected theoretical result. By comparing numerical results with
theoretical values we have verified that the KLA approach is not effective.
Although we do not have any irrefutable reasoning which explains the results
reported herein above, we believe that they are a demonstration of the bias
introduced by the $Q$ entropic index in the weight the probability $p\left(
x\right)  $ in Eq.~(\ref{s-q})~[\cite{ct-bjp}]. Explicitly, for values of $Q>1$
($q<1$) we have $\left[  p\left(  x\right)  \right]  ^{Q}>p\left(  x\right)  $
if $p\left(  x\right)  >1$ and $\left[  p\left(  x\right)  \right]
^{Q}<p\left(  x\right)  $ otherwise. On the other hand, if $Q<1$ ($q>1$) we
have $\left[  p\left(  x\right)  \right]  ^{Q}<p\left(  x\right)  $ if
$p\left(  x\right)  >1$ and $\left[  p\left(  x\right)  \right]  ^{Q}>p\left(
x\right)  $ if $p\left(  x\right)  <1$. Apparently, this bias is overestimated
for $q<1$ and underestimated for $q>1$ by the evaluation of the $\left\langle
\ln_{q}\,\delta\right\rangle $. In the case of uniform PDF, the bias is shed and the KLA yields
a remarkable result.
For $q=1$, the accuracy of the algorithm only diminishes when dependent time series are analysed.

Regarding the Renyi entropic form we have mentioned,
$S_{R} =(\ln \int\left[  p\left(  x\right)  \right]  ^{\alpha }\,dx)/(1 - \alpha )$, ($\alpha \geq 0$),
a similar approach can be implemented, albeit a description involving averages similar to Eqs.~(\ref{2-q_surprise}) and (\ref{sigma-estimator})
is non-trivial. Nonetheless, allowing for the fact that at the first order $S_R = S_Q$ ($\alpha = Q$), further work should deem whether
the remaining terms in the expansion of $S_R$ either set off the error presented by the first approximation (leading to the effectiveness
of the KLA) or sum up to it.

Overall, bearing in mind its importance for a
reliable study of many complex phenomena, it is expected that new binless or binning strategies~[\cite{swinney}] for the
evaluation of entropic functionals such as $S_{Q}$ will correct the shortcoming
conveyed here by the KLA approach.

\begin{acknowledgments}
{\small The work herein presented has benefited from the Marie Curie Fellowship
Programme (European Union).}
\end{acknowledgments}

\appendix

\section{Linearly dependent case}

For this case the variables were obtained by Euler implementing the following
stochastic differential equation~[\cite{borland-pre}],
\begin{equation}
dx=-\gamma \, x\,dt+\sqrt{\theta }\left[ p\left( x,t\right) \right] ^{\frac{\nu }{2}%
}dW_{t}.  \label{dinamica-estavel}
\end{equation}%
The probability density function, $p\left( x,t\right) $, is obtained from the
following non-linear Fokker-Planck equation,
\begin{equation}
\frac{\partial p(x,t)}{\partial t}=\frac{\partial }{\partial x}\left[
\gamma \,x\,p(x,t)\right] +\frac{1}{2}\frac{\partial ^{2}}{\partial x^{2}}\left\{
\theta \,\left[ p\left( x,t\right) \right] ^{\left( 1+ \nu \right) }\right\} ,
\label{fokker-planck-estavel}
\end{equation}%
the solution of which is~[\cite{ct-bukman-1996}],
\begin{equation}
p\left( x,t\right) =\frac{1}{Z_{q}\left( t\right) }\exp _{q}\left[ -\beta
\left( t\right) \,x^{2}\right] ,  \label{q-gauss0}
\end{equation}%
where $q=1- \nu $ and
\begin{equation}
\frac{\beta \left( t\right) }{\beta \left( t_{0}\right) }=\left[ \frac{%
Z_{q}\left( t_{0}\right) }{Z_{q}\left( t\right) }\right] ^{2},
\end{equation}%
\begin{equation}
Z_{q}\left( t\right) =Z_{q}\left( t_{0}\right) \left[ \left( 1-\frac{1}{%
\mathcal{K}}\right) e^{- \gamma \, t }+\frac{1}{\mathcal{K}}\right] ^{1/\left(
2 + \nu \right) },
\end{equation}%
and%
\begin{equation}
\mathcal{K=}\frac{\gamma \, \left[ Z_{q}\left( t_{0}\right) \right] ^{\nu }}{\beta \left( t_{0}\right) \,\theta \,\left( 1+ \nu \right) %
}.
\end{equation}%
The relaxation of the normalisation constant, $Z_{q}$, occurs with the
characteristic time $\tau $,%
\begin{equation}
\tau =\frac{\gamma }{\left( 2 + \nu \right) },
\end{equation}%
which is of the order of $\gamma ^{-1}$ and $-2< \nu <1$ so that $p(x,t)$ is
normalisable. All the correlations for this process are due t the drift term
and because of that the correlations are exponentially decaying. The form of eq. (\ref%
{dinamica-estavel}) corresponds to an equation in which variance is not
constant. The time dependence of the variance leads to the emergence of an
asymptotic power-law behaviour for the probability density function~[\cite{gardiner}].

When $\gamma$ is positive and $\gamma (t-t_{0})\ll 1$, $p(x,t)$ is infinitesimally distant from the stationary
solution~(\ref{fokker-planck-estavel}),%
\begin{equation}
p_{s}\left( r\right) =\frac{1}{Z}\exp _{q}\left[ -\frac{\gamma \,Z^{\nu }}{\left( 1+\nu \right)
\,\theta }x^{2}\right] ,  \label{p-estacionario-estavel}
\end{equation}%
where%
\begin{equation}
Z=\left\{ \frac{\sqrt{2\pi }\,\Gamma \left[ - \frac{1}{\nu}\right] }{\Gamma %
\left[ -\frac{2+ \nu}{2\, \nu}\right] }\sqrt{-\frac{\left( 1+\nu \right) \theta }{%
2\gamma \, \nu }}\right\} ^{\frac{2}{2+\nu}}.
\end{equation}
It is worthless saying that for $\nu = 0$ eq.~(\ref{dinamica-estavel}) recovers the well-known Ornstein-Ulhenbeck
diffusion process~[\cite{feller1}].

\begin{figure}[tbh]
\begin{center}
\includegraphics[width=0.6\columnwidth,angle=0]{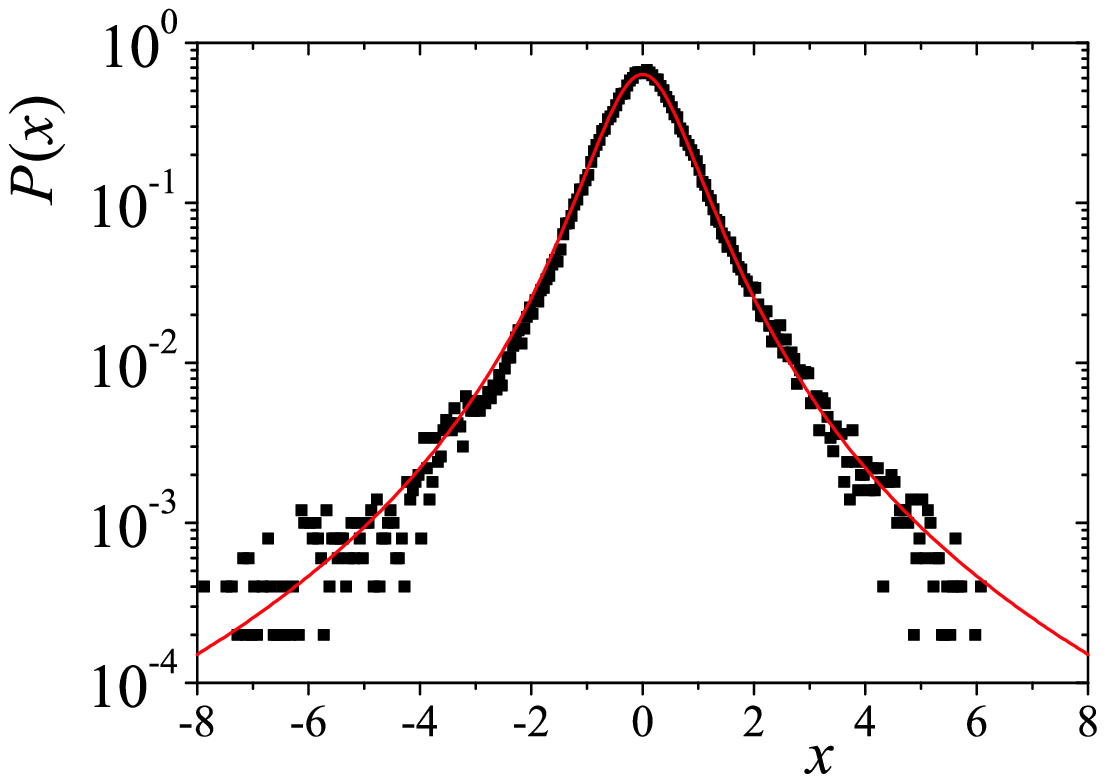}
\end{center}
\caption{ The symbols are casted from a time series of $10^5$ elements generated according to
eq.~(\ref{dinamica-estavel})} and the values mentioned in the text while the line represents the curve
that depicts the $q=3/2$-Gaussian~\ref{q-gauss0}.%
\label{fig-1e}%
\end{figure}

In Fig.~\ref{fig-1e} we plot the correlation function and the histogram of a typical process with
parameters $\gamma = 100^{-1}$, $\theta = \gamma \sqrt{2 / \pi }$ and $\nu = -1/2$.

\section{Non-linearly dependent model}

\begin{figure}[tbh]
\begin{center}
\includegraphics[width=0.6\columnwidth,angle=0]{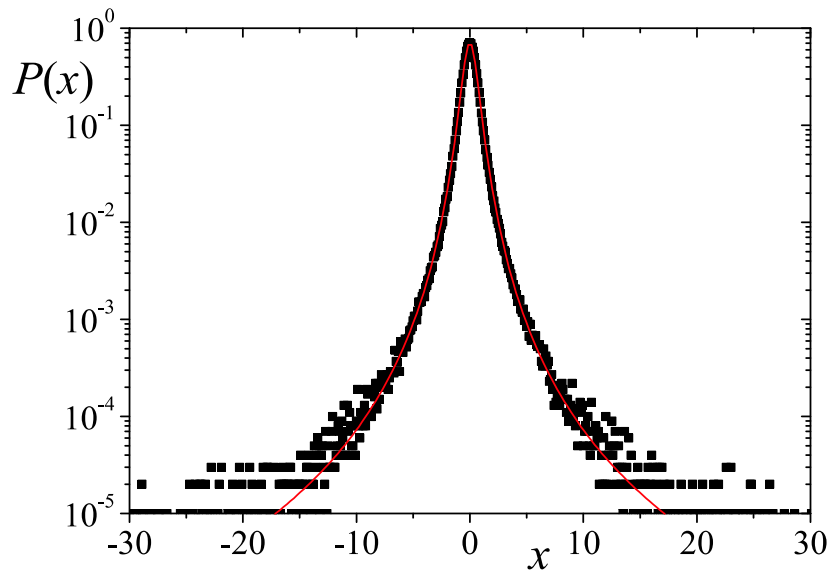}
\end{center}
\caption{ The symbols are casted from a time series of $10^6$ elements generated according to
generalised ARCH process with the parameters mentioned in the text while the line represents the curve
that depicts the $(q=1.54)$-Gaussian with unitary standard variation which fails to rejects the null hypothesis for $\alpha = 0.05$ considering the
Yates' $\chi ^2$-test~[\cite{yates}] and $R ^2 = 0.9997$.}%
\label{fig-2e}%
\end{figure}

In respect of non-linearly dependent models we have opted to consider a modification celebrates
$ARCH$ processes in which an effective immediate past return, $\tilde{x}_{t-1}$, is assumed
in the evaluation of the (squared) volatility $\sigma _{t}^{2}$~[\cite{q-archepl}]. Explicitly,
\begin{equation}
\sigma _{t}^{2}=a+b\,\tilde{x}_{t-1}^{2},\qquad \left( a,b\geq 0\right) ,
\label{vol-qarch}
\end{equation}%
where the effective past return is calculated according to
\begin{equation}
\tilde{x}_{t-1}^{2}=\sum\limits_{i=t_{0}}^{t-1}\mathcal{K}\left(
i-t+1\right) \,x_{i}^{2},  \label{zefect}
\end{equation}%
with
\begin{equation}
\mathcal{K}\left( t^{\prime }\right) =\frac{1}{\mathcal{Z}_{\zeta }\left(
t^{\prime }\right) }\exp _{\zeta }\left[ \frac{t^{\prime }}{T}\right] ,\qquad
\left( t^{\prime }\leq 0,T>0,q_{m}<2\right) ,  \label{kernel}
\end{equation}%
and $\mathcal{Z}_{\zeta }\left( t^{\prime }\right) \equiv \sum_{i=-t^{\prime
}}^{0}\exp _{\zeta }\left[ \frac{i}{T}\right] $. This proposal can be
enclosed in the fractionally integrated class of heteroskedastic process ($FIARCH$)~[\cite{fiarch}].
Although it is similar to other proposals, it has a
simpler structure which permits some analytical considerations without
introducing any underperformance when used for mimicry proposes.
For $\zeta =-\infty $, we obtain the regular $ARCH\left( 1\right) $.
Assuming stationarity in the process the covariance $\left\langle x_{t}^{2}\,x_{t^{\prime }}^{2}\right\rangle $ presents a $q_{c}$%
-exponential form
\begin{equation}
\left\langle x_{t}^{2}\,x_{t+\tau }^{2}\right\rangle \underset{t\rightarrow
\infty }{\sim }\exp _{q_{c}}\left[ -\lambda \,\tau \right] ,\qquad \left(
\tau \geq 0\right)  \label{c2}
\end{equation}%
with
\begin{equation}
q_{c}=\frac{1}{2-\zeta },  \label{qc-qm}
\end{equation}%
and $\lambda =q_{c}^{-1}$.
This long-lasting correlation of the volatility comes forth in the the shape of non-linear correlations in $x$ that
are gauged by Kullback-Leibler related measures~[\cite{q-archepjb}]. Like the standard form of the ARCH process,
the analytical expression of the stationary probability density function remains unknown. Nonetheless, there is
robust statistical evidence that it is well-described by the $q$-Gaussian form.
For the value used in the article $\zeta = 1.375$, $b = 0.9375$ and $a=1-b$ we have verified that the distribution is well fitted by
a $(q=1.54)$-Gaussian with unitary standard deviation, which agrees with the value found, e.g., for the daily index fluctuations of the Dow Jones 30. In Fig.~(\ref{fig-2e}) we plot the histogram of a heteroskedastic process with the parameters aforementioned.

\begin{thebibliography}{99}
%
\bibitem[Abramov {\it et al.}(2007)RA] {abramov}R. Abramov, J. Comp. Phys. \textbf{226}, 621 (2007); R.
Abramov, J. Comp. Phys. \textbf{228}, 96 (2009)

\bibitem[Andersen {\it et al.}(2005)TGA]{fiarch} T.G. Andersen, T. Bollerslev, P.F. Christoffersen, F.X.
Diebold, \textit{Volatility Forecasting}, PIER working paper 05-011 (2005)

\bibitem[Bailey(1994)RWB] {bailey}R.W. Bailey, \textit{Polar Generation of Random Variates with
the }$t$\textit{-Distribution}, Math. of Comp. \textbf{62}, 141 (1994)

\bibitem[Borland(1998)]{borland-pre} L. Borland, Phys. Rev. E \textbf{57}, 6634 (1998)

\bibitem[Caruso \& Tsallis(2008)]{ct-caruso} F. Caruso and C. Tsallis, Phys. Rev. E \textbf{78}, 021102 (2008)
Sci. Comp. \textbf{26}, 411 (2003)

\bibitem[Cohen(1996)]{cohen} E.G.D. Cohen, Boltzmann and Statistical Mechanics,
in:\ \textit{Proceedings of the International Meeting Boltzmann's Legacy 150
Years After His Birth}, Atti della Accademia Nazionale dei Lincei, 9-23 (1997). Also at \texttt{arXiv:cond-mat/9608054v2} (preprint, 1996)

\bibitem[Engle(1982)]{engle} R.F. Engle, Econometrica \textbf{50,} 987 (1982)

\bibitem[Fraser \& Swinney(2002)] {swinney} A.M. Fraser and H.L. Swinney, Phys. Rev. E \textbf{66}, 2002

\bibitem[Feller(1971)]{feller1} W. Feller, \textit{An Introduction to Probability Theory
and its Applications} (Whiley, New York, 1971)

\bibitem[Gardiner(2004)]{gardiner} C.W. Gardiner, \textit{Handbook of Stochastic Methods for
Physics, Chemistry and the Natural Sciences} Third Edition (Springer-Verlag,
Berlin, 2004)

\bibitem[Gentle(1995)]{gentle}J.E. Gentle, \textit{Random Number Generation and Monte Carlo
Methods}, vol.2 2nd ed. (John Wiley \& Sons, New York, 1995)

\bibitem[Gradshteyn \& Ryzhik(1980)] {gradshteyn} I.S. Gradshteyn and I.M. Ryzhik, Table of Integrals,
Series and Products (Academic Press, New York, 1980) \texttt{3.191.3}

\bibitem[Huang(1963)] {huang}K. Huang, Statistical Mechanics (John Wiley \& Sons, New York, 1963)

\bibitem[KLA(1987)]{kla} L.F. Kozachenko and N.N. Leonenko, Prob. Inf. Trans.
\textbf{23}, 95 (1987); P. Grassberger, Phys. Lett. A \textbf{107}, 101
(1985); J.D. Victor, Phys. Rev. E \textbf{66}, 051903 (2002)041904 (2002)

\bibitem[Kraskov {\it et al.}(2004)]{klakl} A. Kraskov, H. Stoegbauer and P. Grassberger, Phys. Rev. E
\textbf{69}, 066138 (2004)

\bibitem[Renyi(1970)] {renyi} A. Renyi, Probability Theory (North-Holland, Amsterdam, 1970)

\bibitem[Risken(1989)]{risken} H. Risken, \textit{The Fokker-Planck Equation: Methods of
Solution and Applications}, $2^{nd}$ edition (Springer-Verlag, Berlin, 1989)

\bibitem[SMDQ(2007)]{q-archepl} S.M. Duarte Queir\'{o}s, EPL \textbf{80}, 30005 (2007)

\bibitem[SMDQ(2008)]{q-archepjb} S.M.D. Queir\'{o}s, Eur. Phys. J. B \textbf{66}, 137 (2008)

\bibitem[SMDQ(2009)] {phys-d} S.M. Duarte Queir\'{o}s, Physica D \textbf{238}, 764 (2009)

\bibitem[Souza \& Tsallis(1997)]{andre} A.M.C. de Souza and C. Tsallis, Phys. A \textbf{236}, 52
(1997)

\bibitem[Tsallis(1988)] {tsallis} C. Tsallis, J. Stat. Phys. \textbf{52}, 479 (1988)

\bibitem[Tsallis \& Bukman(1996)]{ct-bukman-1996} C. Tsallis, D.J. Bukman, Phys. Rev. E \textbf{54},
R2197 (1996)

\bibitem[Tsallis(1999)] {ct-bjp} C. Tsallis, Braz. J. Phys. \textbf{29}, 1 (1999)

\bibitem[Tsallis {\it et al.}(2005)]{tsallis-pnas} C. Tsallis, M. Gell-Mann and Y. Sato, Proc. Nat. Acad.

\bibitem[Tsallis(2009)]{book} C. Tsallis, \textit{Introduction to Nonextensive Statistical
MEchanics: Approaching a Complex World} (Springer, Berlin, 2009). A
comprehensive list of applications in widespread fields of science and
knowledge is available at \texttt{http://tsallis.cat.cbpf.br/biblio.htm}Sci. USA \textbf{101}, 15852 (2005)

\bibitem[Volkovski \& Sinai(1971)]{sinai} K.L. Volkovyski and Ya.G. Sinai, Funct. Anal. Appl.
\textbf{5}, 185 (1971)

\bibitem[Yates(1934)]{yates} F. Yates, J. Royal Stat. Soc. \textbf{1}, 217 (1934)

\end{thebibliography}
\end{document}